\documentstyle[12pt,aaspp4]{article} %preprint format

\lefthead{OHYAMA ET AL.}
\righthead{SUBARU SPECTROSCOPY OF HST~14176$+$5226}

\begin{document}

\title{Subaru Spectroscopy of the Gravitational Lens HST~14176$+$5226:
Implications for a Large Cosmological Constant\footnote{
Based on observations made with the Subaru Telescope, which is
operated by the National Astronomical Observatory of Japan.}}

\author{Youichi Ohyama,}
\affil{Subaru Telescope, National Astronomical Observatory of Japan,
650 N. A`ohoku Place, University Park, Hilo, HI 96720}
\author{Takashi Hamana, Nobunari Kashikawa, Masashi Chiba,}
\affil{National Astronomical Observatory of Japan, Mitaka, Tokyo 181-8588,
Japan}
\author{Toshifumi Futamase}
\affil{Astronomical Institute, Tohoku University, Sendai 980-8578, Japan}
\author{Masanori Iye, Koji S. Kawabata}
\affil{National Astronomical Observatory of Japan, Mitaka, Tokyo 181-8588,
Japan}
\author{Kentaro Aoki,}
\affil{Institute of Astronomy, University of Tokyo, Mitaka, Tokyo 181-0015,
Japan}
\author{Toshiyuki Sasaki, George Kosugi, and Tadafumi Takata}
\affil{Subaru Telescope, National Astronomical Observatory of Japan,
650 N. A`ohoku Place, University Park, Hilo, HI 96720}

\begin{abstract}
We present new optical spectroscopy of the lens elliptical galaxy
in the ``Einstein Cross'' lens system HST~14176$+$5226, using
the Faint Object Camera and Spectrograph (FOCAS) of the Subaru telescope.
Our spectroscopic observations are aimed at measuring the stellar velocity
dispersion of the lens galaxy, located at high redshift of $z_L=0.81$,
as an important component to lens models. We have measured this dispersion to be
$230 \pm 14$ km~s$^{-1}$ (1~$\sigma$) inside 0.35 effective radii of the lens,
based on the comparison between the observed galaxy spectrum and spectral
templates of three G-K giants by means of the Fourier cross-correlation method.
To extract the significance of this information on the geometry of the
universe which also affects the lensing of the background image, we attempt
to fit three different lens models to the available data of the lens system.
Provided that the lens galaxy has the structural and dynamical properties
(i.e., its radial density profile, core radius, and velocity anisotropy)
similar to those of local elliptical galaxies, we calculate the likelihood
function for the simultaneous reproduction of both the observed image splitting
and newly measured velocity dispersion of the lens. Although the confidence
interval depends rather sensitively on the adopted lens models or their
parameters, our experiments suggest the larger likelihood for a larger
cosmological constant, $\Omega_\Lambda$: formal 1~$\sigma$ lower limit on
$\Omega_\Lambda$ in the flat universe ranges 0.73 to 0.97, whereas 2~$\sigma$
lower limit is basically unavailable. This method for determining the world
model is thus dependent on lens models but is insensitive to other unavoidable
ambiguities, such as the dust absorption or the evolutionary effects of
galaxies. Exploring spectroscopic observations of more lens galaxies at high
redshift may minimize the model uncertainties and thus place a much tighter
constraint on $\Omega_\Lambda$.
\end{abstract}

\keywords{cosmology: observations -- gravitational lensing
--- quasars: individual (HST~14176$+$5226)}

%%%%% Sec. 1 %%%%%
\section{Introduction}

Recent cosmological observations have increased various lines of evidence
that the evolution of the universe may be dominated by a nonvanishing,
normalized cosmological constant,
$\Omega_\Lambda$ ($\equiv \Lambda c^2/3H_0^2$, where
$H_0$ is a Hubble constant). These include the distance determination of
high-redshift Type Ia supernovae (SNe) (e.g. Perlmutter et al. 1998),
number counts of faint galaxies (Fukugita et al. 1990; Yoshii \& Peterson 1995),
statistics of gravitationally lensed QSOs (Chiba \& Yoshii 1999), and
age calibration of the Galactic globular clusters (Chaboyer et al. 1998).
Also, recent measurements of small-scale anisotropy in the cosmic microwave
background (CMB) radiation suggest the flat geometry of the universe
($\Omega_m+\Omega_\Lambda=1$, where $\Omega_m$ stands for the matter density),
which may be reconciled with most of the inflationary universe scenarios
(de Bernardis et al. 2000). If this result is combined with the currently
favored value of $\Omega_m=0.2 \sim 0.3$ (Bahcall et al. 1999), then
a large $\Omega_\Lambda$ ranging 0.7 to 0.8 is inferred.

Except for the CMB measurements, all of the astronomical observations
to determine $\Omega_\Lambda$ suffer from unavoidable ambiguities associated
with dust absorption inside and/or outside the observed objects in concern,
their evolutionary effects, and selection effects. For instance, the
interpretation of the Type Ia SNe results should be modified non-trivially if
their luminosities are affected by dust in their host galaxies or intergalactic
medium (Aguirre 1999; Totani \& Kobayashi 1999) or if their intrinsic
properties at high redshift differ from local counterparts (Riess et al. 1999,
but see Aldering et al. 2000).
Although continuing efforts to minimize these sources of ambiguities have
been progressively made, especially by observing higher-redshift SNe
(Riess et al. 2001), it may be equally worth pursuing an alternative,
independent methodology for the determination of $\Omega_\Lambda$,
which exempts from the above ambiguities.

In this regard, gravitational lensing may be the most powerful and accurate
probe to measure cosmological parameters, because the physics involved in it
is gravity only (e.g. Mellier 1999). Among various proposed methods relied on
gravitational lensing, an interesting approach was adopted by Im et al. (1997),
using the dependence of strong lenses on $\Omega_\Lambda$
(Paczy\'nski \& Gorski 1981; Gott et al. 1989).
They investigated the mean splitting of
the lensed multiple images of a distant QSO, $\langle\Delta\theta\rangle$, in
conjunction with the mass of a foreground lens galaxy as deduced from its
line-of-sight velocity dispersion, $\sigma$. They deduced $\sigma$ from
the observed lens magnitude, lens redshift, and the Faber-Jackson relation
after correcting for luminosity evolution of a lens. If the mass distribution
of a lens is simply represented by a singular isothermal sphere (SIS) as
they adopted, we obtain,
\begin{equation}
\langle\Delta\theta\rangle = 8\pi \left( \frac{\sigma}{c} \right)^2
                       \frac{D_{ls}}{D_s} ,
\end{equation}
where $D_s$ and $D_{ls}$ denote angular diameter distances to the source
and between the lens and source, respectively, both of which depend on
lens redshift $z_L$, source redshift $z_S$, and cosmological parameters
$(\Omega_m,\Omega_\Lambda)$. Applying this method to eight lens systems,
Im et al. concluded $\Omega_\Lambda = 0.64_{-0.26}^{+0.15}$ (1~$\sigma$)
in the flat
universe, and $\Omega_m=1$ is excluded at the 97~\% confidence level.

Im et al.'s procedure is subject to ambiguities in dust absorption
(Keeton et al. 1998) and in evolution correction for the lens galaxies.
Also, the use of the simple SIS model appears to be too simplified.
However, as equation (1) implies, if the information on velocity dispersion
of a lens is available from {\it direct} spectroscopic observation,
then the concerned ambiguities of dust absorption and galaxy evolution
can be excluded in the determination of $\Omega_\Lambda$.
Furthermore, if a more realistic lens model than SIS is applied and
if a desired combination of $(z_L,z_S)$ for the lens system is chosen
so as to make the distance ratio $D_{ls}/D_s$ being sensitive to
$\Omega_\Lambda$, then we may be able to place a more reliable limit
on $\Omega_\Lambda$. To highlight the latter point, we plot,
in Figure 1, the relation between $\Omega_m$ and the predicted $\sigma$
from equation (1), for the case of $z_L=0.81$, $z_S=3.40$, and
$\langle\Delta\theta\rangle/2=1.\arcsec5$ (solid line) approximately
corresponding to the lens
system HST~14176$+$5226 employed in this paper, and for the case of
$z_L=0.30$ while other quantities are fixed (dotted line). For each case,
the upper and lower lines correspond to $\Omega_\Lambda=0$ and
$\Omega_m+\Omega_\Lambda=1$, respectively. This plot clearly demonstrates that
the effect of $\Omega_\Lambda$ on $\langle\Delta\theta\rangle$ can
be more easily
separated from that of the lens mass and/or uncertainties in the mass
distribution, if more remote lenses, say at $z_L \sim 1$, are utilized
in this analysis.
A similar method using Einstein ring systems has been proposed by
Futamase and Hamana (1999).

Based on the above motivation, we conducted the direct measurement
of velocity dispersion for the lens galaxy of the ``Einstein Cross''
HST~14176$+$5226, using the FOCAS (Kashikawa et al. 2000) mounted
on the Subaru telescope. This lens system was serendipitously discovered
in the WFPC2 images from the {\it HST} Medium Deep Survey
(Ratnatunga et al. 1995; 1999; Knudson et al. 2001).
The lensed source appears to be a QSO at $z_S=3.4$ (Crampton et al. 1996;
Moustakas \& Davis 1998), splitted into four images in a symmetric
configuration. The foreground lens galaxy is a red, remote
elliptical galaxy at $z_L=0.81$, and its bright apparent magnitude
(F814W$=19.8$, F606W$=21.7$) enabled us to obtain adequate spectra of 
high enough S$/$N with the 8.2~m telescope.
Thus, combined the observational results with the detailed lens
modeling, it may be possible to place a useful limit on $\Omega_\Lambda$.

This paper is organized as follows. In \S 2, we present the spectroscopic
observations and data reduction. Our technique to calibrate the line-of-sight
velocity dispersion from the derived spectrum is described in \S 3. In \S 4,
we describe the lens modeling and stellar dynamics of the lens galaxy.
Model fittings to the lens system and the results for the determination
of $\Omega_\Lambda$ are presented. In \S 5, implications of the results
and further prospects of the work are discussed.

%%%%% Sec. 2 %%%%%
\section{Observations and Data Reduction}

Observations were made with the FOCAS attached on the Subaru 8.2 m telescope
on 2001 June 18. The seeing condition was 0\arcsec .8-0\arcsec .9 during the
observation. We used the narrow 0\arcsec .4 slit along the major axis of the
lens galaxy (PA$=-41.5^\circ$). The narrow slit was selected aiming for higher
wavelength resolution for better velocity measurement accuracy. The $300$
lines mm$^{-1}$ grating together with a Y47 order-cut filter allowed us to
obtain an optical spectrum between 4700\AA~ and 9000\AA~ with a pixel
resolution of 1.39\AA. The spatial resolution was 0\arcsec .4 per pixel by
using 4-pixel on-chip binning. We obtained six 1800 seconds exposures.

After subtracting the bias in a standard manner, flat fielding, the optical
distortion corrections, and cosmic ray removal were applied using the dedicated
software (Yoshida et al. 2000) for each frame. Wavelength calibration was made
based on both the calibration Th-Ar lamp and the telluric OH lines. The
accuracy of the wavelength calibration was measured with OH lines to be
$\simeq 0.12$\AA~ in RMS over 5500\AA~ -- 9000\AA. Flux calibration was made
with a spectroscopic standard star Wolf 1346 and the atmospheric extinction
was corrected by using the standard Mauna-Kea extinction curve with a standard
manner using IRAF
\footnote{IRAF is distributed by the National Optical Astronomy Observatories,
     which are operated by the Association of Universities for Research
     in Astronomy, Inc., under cooperative agreement with the National
     Science Foundation.}.
The lens galaxy spectra were extracted through a 0\arcsec .8 aperture around
the nucleus (a 2-pixel wide aperture around the brightest column) from each
frame to yield six 1-dimensional spectra.
This aperture corresponds to
$2 \times 0.35 R_e$, where $R_e (=0.\arcsec13 \pm 0.\arcsec 05)$ denotes an
effective radius of a de Vaucouleurs law for the surface brightness profile
(Keeton et al. 1998). Finally they were combined with 3-$\sigma$ clipping
average algorithm to remove the residual cosmic ray features. Following
van Dokkum and Franx (1996), the instrumental spectral resolution was
measured using several OH lines. For this purpose, the sky spectrum was made
following the same procedure as for the galaxy spectrum (i.e., using the same
wavelength transformation, the aperture extraction, and the frame combination
methods).We selected six narrowest (unblended) OH lines around 7300\AA~
where our target absorption lines of lens galaxy reside and they were fitted
simultaneously with multiple Gaussian components whose line widths were
assumed to be identical for all lines. We found the instrumental resolution
$5.6\pm 0.3$\AA~ in full width at half maximum (FWHM).

The spectrum of the lens galaxy of HST~14176$+$5226 is shown in Figure 2. It
shows deep CaII H and K absorption lines (3970\AA~ and 3932\AA~ detected at
7120\AA~ and 7185\AA~, respectively) and prominent G band feature (4399\AA~
detected at 7800\AA), suggesting that major contribution in the observed flux
of this spectral region could be attributed to the late type giant stars
(late G giant $\sim$ early K giant stars).

%%%%% Sec. 3 %%%%%
\section{Velocity Dispersion Measurement}

The Fourier cross-correlation method (Tonry \& Davis 1979) was used to measure
the line-of-sight velocity dispersion of the lens galaxy with the FXCOR task
implemented in IRAF. This method compares the stellar spectrum as a template
with the observed galaxy spectrum by means of the cross-correlation technique.
Because we cannot analyze the stellar population in detail given the smaller
wavelength coverage and the limited quality of the spectrum, we assume that
a single representative population of stars can reproduce the galaxy spectrum
well. Therefore we selected three late type giant stars [HD 83805 (G8III),
HD 8491 (K0III), and HD 94247 (K3III)] as spectral templates. All templates
were used in the FXCOR task separately to estimate the $\sigma$ uncertainties
in regard to the uncertainties of types of the template. Detailed properties
of the template stars used are given in Table 1. The spectra were taken from
the stellar spectra library of the C\'oude feed spectral library published
in the AAD Volume 7 CD-ROM (Leitherer et al. 1996). They cover 3800-4900\AA~
with a wavelength resolution of 1.8\AA~ (FWHM), and are suitable in our
analysis because the resolution is higher than that of the instrumental
resolution at the rest-frame of the lens galaxy (3.1\AA~ FWHM) and all
the prominent redshifted features detected in the lens galaxy spectrum
(such as Ca H+K and G band) are included in the templates.

Since the velocity resolution matching between the galaxy and the template
spectra is crucial to obtain proper velocity dispersion (e.g. Falco et al.
1997), convolution with the Gaussian-broadening function was applied to the
templates (53 km s$^{-1}$ resolution at the G band) to make the velocity
resolution-matched templates (the rest-frame resolution of the lens galaxy
spectrum: $91\pm 5$ km s$^{-1}$).
We estimate that the error in the velocity resolution matching would result in
the error of the velocity dispersion of 4 km s$^{-1}$ ($1\sigma$):
The solution of the velocity dispersion of the lens galaxy ($\sigma$) in
the Fourier correlation method is related to the instrumental resolution
($\tau$) as $\sigma^2= \mu^2 - 2\tau^2$ where $\mu$ is the width of the cross
correlation peak (Tonry \& Davis 1979), and, hence, $\delta \sigma \simeq 2\tau \delta \tau / \sigma \simeq 4$ km s$^{-1}$ for $\sigma = 230$ km s$^{-1}$ (see below), $\tau = 91$ km s$^{-1}$, and $\delta \tau = 5$ km s$^{-1}$.

Following Falco et al. (1997), we analyzed the velocity dispersion of the lens
galaxy in the following ways. First, the spectra were pre-processed to be used
in FXCOR. Spectra around 7050\AA - 8200\AA~ for the lens galaxy, or
3900\AA~ - 4500\AA~ for templates, were selected to include interesting
absorption features such as Ca H, K and G band. All spectra were normalized by
the continuum which was obtained by fitting the spectra with the low-order
polynomial function. In this fitting, strong absorption features such as
Ca H and K, G band, and atmospheric A band were excluded. Using these spectra,
we measured the redshift of the galaxy as $z = 0.811$ by the FXCOR. Note that
the redshift was determined consistently regardless of the stellar templates
employed and of the detailed FXCOR parameters due to strong absorption
features. Our redshift estimate is almost consistent with Koo et al.
(1996, $z=0.811$) and Crampton et al. (1996, $z=0.809$).
Adopting thus derived redshift of the lens galaxy, we produced the rest-frame
lens galaxy spectrum (blue-shifted to $z=0$) for further line width analysis.

Then we made calibration curves which relate the velocity dispersion
($\sigma$) with the FWHM of the cross-correlation function (CCF) peak measured
with the FXCOR (see Falco et al. 1997). Resolution-matched templates were
convolved with a series of Gaussian-broadening functions in which the velocity
dispersion ranges from 45 km s$^{-1}$ to 270 km s$^{-1}$. One can obtain
the calibration curve by running the FXCOR for various combinations of
the resolution-matched templates and the Gaussian-broadened templates.
In FXCOR, we used the following parameters to calculate the CCF and to
measure the width of its peak. We tried two spectral regions for each template
(two out of the following three regions: A: $4250-4450$\AA, B: $4250-4470$\AA,
and C: $4250-4500$\AA) in calculating the CCF. The regions were selected to
include G band but to exclude the atmospheric A band and Ca H and K lines
since the calculation with Ca H and K would give problematic result because of
their intrinsically wider line width (Tonry 1998). Apodization of 5\% (at each
end) was applied on all spectra before the Fourier transformation to minimize
the spectral aliasing effects. Lowest wavenumber portion of
the Fourier-transformed spectra ($\lesssim 110$\AA$^{-1}$) were not used to
avoid the effect of the continuum variation remaining even after the continuum
normalization procedures. After the cross-correlation, the FWHM of the CCF
peak were measured by Gaussian profile fitting over either central 30 or 35
lags ($\sim \pm 674$ or $\pm 786$ km s$^{-1}$ from the peak) around the peak.
The reason why we used two fitting wavelength regions on spectra and two
fitting widths on CCF is to estimate the $\sigma$ uncertainties in the FXCOR
fitting procedure. In this way, we got 12 calibration curves in total (three
templates, two wavelength fitting regions per template, and two lag widths per
each combination of the templates and the wavelength fitting region).

Next, we ran the FXCOR twelve times with the rest-frame lens galaxy spectrum
and the resolution-matched templates to measure the FWHM of the CCF peak.
The same parameters of the FXCOR as in making the calibration curves were used
in each run.
The peaks of the CCFs could be represented well by a Gaussian function like
in making calibration curves, and the fittings were made with a typical error
of $\simeq 10$ km s$^{-1}$ for all cases.
In order to check how well the CCF peak fitting could be made and to know how
the noise signal of the observed galaxy spectrum could possibly affect
the measurement, thirty artificial noise spectra were made with the mknoise
task of the artdata package in IRAF, and were added to the observed galaxy
spectrum to increase the root-mean-square noise level by $\sim 20$\%.
Then, we ran the FXCOR thirty times with the noise-added spectra for one case
of the calibration curve, and found that a scatter of the results is
$\simeq 12$ km s$^{-1}$ with the mean value being almost as same as that for
the original spectrum within 1 km s$^{-1}$.
Note that, although the scatter is slightly worse than the fitting error
for one CCF width measurement with the original spectrum (without artificial
noise), the small increase ($\sim 20$\%) is most likely to come from
the artificial noise added to the observed spectrum.
Considering all these experiments, we adopted a 1$\sigma$ uncertainty of
10 km s$^{-1}$ in the CCF measurement.

The twelve CCF peak width values obtained with original lens galaxy spectrum
for various parameters were then converted to velocity dispersion by using
the corresponding calibration curves. We
obtained $\sigma = 217 \sim 244$ km s$^{-1}$ with moderately good Tonry-Davis
$R$ values ($R=12-15$) for all FXCOR runs (Table 2). Here the Tonry-Davis $R$
value gives the signal-to-noise ratio of the CCF peak (Tonry and Davis 1979),
and is also shown to indicate how well the galaxy and template spectra are
correlated with each other.
Since $\sigma$ obtained for each calibration curve
are almost similar, we adopt the $R$-weighted mean $\sigma$ value of
230 km s$^{-1}$ as a most likely value, and the $1\sigma$ uncertainties
originating from the choice of different templates, fitting wavelength
regions, and CCF fitting widths would be $\sim 8$ km s$^{-1}$ based on the
scatter of the derived $\sigma$ for each FXCOR run.
Figure 3 shows the normalized rest-frame lens galaxy spectrum overlaied by
the Gaussian-broadened ($\sigma = 230$ km s$^{-1}$) template of HD 83805
(the case of the best $R$ value) as well as the residual spectrum (galaxy
spectrum $-$ Gaussian-broadened template). To show how remaining sky spectrum
may possibly affect the fitting, we also show below the sky spectrum which is
red-shifted by the same amount as for the lens galaxy spectrum. Wavelength
region around Ca H and K are also shown although the region was not used in
the FXCOR analyses. One may find that the G band profile of the galaxy can be
fitted by the Gaussian-broadened template very well, and Ca H and K lines are
also represented by the template relatively well. Although there are some
remaining features in the residual spectrum such as absorption features at
4000\AA~ and 4435\AA, they are likely to come from the sky subtraction
residual. Therefore, we confirmed that the stellar templates used were indeed
suitable for representing the lens galaxy spectrum.

Finally we summarize errors associated with all above procedures (Table 3).
First, error in the template-galaxy velocity matching is 4 km s$^{-1}$
($1\sigma$) based on the instrumental resolution uncertainty.
The CCF peak width measurement with the velocity-matched spectrum
can be made with 10 km s$^{-1}$ accuracy ($1\sigma$) for one FXCOR
run for a calibration curve whose parameters are spectrum template,
wavelength fitting region, and fitting width of the CCF peak.
Then, twelve FXCOR runs for twelve calibration curves, all of which
seem equally suitable for the FXCOR parameters, give
8 km s$^{-1}$ scatter ($1\sigma$) of the velocity dispersion.
Assuming all these sources of error are independent of each other,
the overall $1\sigma$ uncertainty of the velocity dispersion is estimated
to be $(4^2+10^2+8^2)^{0.5} \simeq 14$ km s$^{-1}$.

%%%%% Sec. 4 %%%%%
\section{Lens Modeling}

\subsection{Lens Models}

In order to demonstrate what constraints the current velocity measurement of
the lens galaxy provides on cosmologies, we investigate lens mass models
which reproduce the key observational information on the lens system
(image positions) as a function
of cosmological parameters. Our lens models are represented by a cored
isothermal ellipsoid (CIE) in an external shear field to take advantage of its
simplicity and also generality. A similar lens model to ours, a singular
isothermal ellipsoid (Kassiola \& Kovner 1993; Kormann et al. 1994), has been
known to be broadly consistent with observations of various lens systems
(e.g. Keeton et al. 1998). An isothermal profile for the total mass
distribution of ellipticals is well supported by the detailed dynamical
studies of local ellipticals (Rix et al. 1997; Gerhard et al. 2001),
individual lens modeling, and statistics
(e.g. Maoz \& Rix 1993; Kochanek 1995; Grogin \& Narayan 1996).
The inclusion of an external shear field appears to be necessary both to
improve the fits of lens models to the data and to make an axis ratio
distribution of individual lenses being consistent with the observed axis
ratio distribution of light (Keeton et al. 1997). We allow a finite central
core in the lens models, because recent {\it HST} photometry for the centers of
ellipticals has revealed the presence of a finite core in bright ellipticals
(Faber et al. 1997; Ravindranath et al. 2001).

The CIE model is parameterized by an ellipticity of mass distribution
$\epsilon$, where the axis ratio is given by
$\sqrt{(1-\epsilon)/(1+\epsilon)}$,
position angle of its semi-major axis $\phi_\epsilon$,
core size $r_c$,
and line-of-sight velocity dispersion $\sigma_{tot}$. Given such a lens,
the lens equation relates the position of a source $(\beta_1, \beta_2)$
on the source plane to the positions of images $(\theta_1^i, \theta_2^i)$
($i = 1$ ... 4 for the four-image case) on the lens plane (where the
positions are relative to the lens center),
\begin{eqnarray}
\beta_1 &=& \theta_1 - \alpha_0
  \frac{(1-\epsilon)\theta_1}{
  \sqrt{(1-\epsilon)\theta_1^2 + (1+\epsilon)\theta_2^2 + \theta_c^2} }
  - \gamma_1 \theta_1-\gamma_2 \theta_2, \\
\beta_2 &=& \theta_2 - \alpha_0
  \frac{(1+\epsilon)\theta_2}{
  \sqrt{(1-\epsilon)\theta_1^2 + (1+\epsilon)\theta_2^2 + \theta_c^2} }
  - \gamma_2 \theta_1 + \gamma_1 \theta_2,
\end{eqnarray}
where $\theta_c$ is the angular size of a core ($=r_c/D_l$, where $D_l$
denotes angular diameter distance to the lens) and  $\alpha_0$
characterizes the strength of a lens as defined by $\alpha_0 \equiv
4\pi (\sigma_{tot}/c)^2 D_{ls}/D_s$. The effects of the external shear,
which is parameterized by the strength
$\gamma=(\gamma_1^2+\gamma_2^2)^{1/2}$ and orientation
$\tan 2 \phi_\gamma=\gamma_2/\gamma_1$, are also included in the above
equations.

We consider three different lens models.
The first is the CIE model without the external shear, which is represented
by ten parameters: four for source and lens
positions, four for lens parameters, and two for cosmological
parameters $(\Omega_m$, $\Omega_\Lambda)$.
The second is the CIE model with the external shear (CIE+ES), which
introduces additional two parameters, so that the total number of parameters
is twelve.
The third is similar to the CIE model but the lens coordinates are treated as
free parameters (CIE+LP).
This takes into account the possible finite difference between the observed
galaxy center and the center of the gravitational potential.
In this model, we do not consider the external shear, so that
the total number of parameters is twelve.

As a part of observational constraints on the models, we adopt the positions of
the four lensed images and the lens galaxy. The flux ratios between the optical
images are excluded in the fitting procedure, because these are largely
affected by dust extinction, microlensing by stellar masses, or dark matter
substructure (e.g. Mao \& Schneider 1998; Chiba 2002). We refer to
{\it HST} data available on the CASTLES Survey's web site
(http://cfa-www.harvard.edu/castles/) for the positions and their uncertainties
($0.\arcsec03$) of the current lens system.  For the positional uncertainty of
the lens galaxy, we adopt a slightly more conservative value ($0.\arcsec05$
instead of $0.\arcsec03$) because of the extended nature of the galaxy image,
although this change of the positional uncertainty turns out {\it not}
to affect the dependence of the image fitting on the model parameters as
performed below.

\subsection{Stellar Dynamics}

The velocity dispersion, $\sigma_{tot}$, as a parameter of the lens model
is not a directly observable quantity, while what we have measured as reported
in the previous section is the surface-brightness weighted average of
the line-of-sight velocity dispersion of a stellar component interior to
a projected radius $R$, hereafter denoted
as $\langle\sigma_{los}^2(R)\rangle^{1/2}$ (Kochanek 1993).
To relate $\langle\sigma_{los}^2(R)\rangle^{1/2}$ with $\sigma_{tot}$,
we assume that the total mass density, consisting of a luminous stellar
and dark matter component, follows a cored isothermal distribution,
to be consistent with our lens model, and that the velocity anisotropy of
stars is measured by a parameter $\beta$, i.e.,
$\sigma_\theta^2 = \sigma_\phi^2 = (1-\beta) \sigma_r^2$ in a spherical
geometry, where $(\sigma_r, \sigma_\theta, \sigma_\phi)$ are velocity
dispersions along the polar coordinates $(r, \theta, \phi)$.
Then, for a luminous stellar component having the surface brightness
$I(R)$ and line-of-sight velocity dispersion $\sigma_{los}(R)$ at a projected
radius $R$, the Jeans equation yields (e.g. Binney \& Tremaine 1987),
\begin{equation}
I(R) \sigma_{los}^2(R) = 2 \int_0^\infty dz
  \left( 1 - \beta \frac{R^2}{r^2} \right) r^{-2\beta}
  \int_r^\infty dr' \frac{\nu(r')GM(r')}{r'^2} r'^{2\beta} ,
\end{equation}
with $r^2=R^2+z^2$, where $\nu(r)$ is the volume luminosity density and
$M(r)$ is the total mass inside $r$, given as $M(r) =
2 \sigma_{tot}^2 r / G (1 - r_c/r \arctan r/r_c)$
(Krauss \& White 1992; Kochanek 1993).
After averaging interior to $R$, we obtain
\begin{equation}
\langle\sigma_{los}^2(R)\rangle = \int_0^R d^2R' I(R') \sigma_{los}^2(R')
  / \int_0^R d^2R' I(R') .
\end{equation}

As is evident from equation (4) and (5), $\langle\sigma_{los}^2(R)\rangle$
depends on two parameters, $r_c$ and $\beta$, in such a way that
$\langle\sigma_{los}^2(R)\rangle$ at a specific radius $R$ increases
with increasing $\beta$ (as demonstrated in Kochanek 1993)
or decreasing $r_c$. To incorporate
these properties into our likelihood analysis described below,
we make a linear regression to the relation between
$\sigma_{tot}^2/\langle\sigma_{los}^2\rangle$ and $r_c$ for a given $R$
and $\beta$,
\begin{equation}
\sigma_{tot}^2 = \langle\sigma_{los}^2\rangle
                  \left( a \frac{r_c}{R_e} + b \right) ,
\end{equation}
where $a$ and $b$ are coefficients and $R_e$ stands for an effective radius
of a de Vaucouleurs law for $I(R)$. We note that the linearity in the relation
is an excellent approximation over the observed range of $r_c$,
i.e., $0 \le r_c/R_e \le 0.05$ (Faber et al. 1997; Ravindranath et al. 2001).
In accordance with our observations reported in Section 2, we set
$R/Re = 0.35$. For $\beta$, we take advantage of the detailed dynamical
studies of local elliptical galaxies (Gerhard et al. 2001). Gerhard et al.
(2001) showed, based on their sample of 21 early-type galaxies, that an average
$\beta$ over the range $0.1R_e$ to $1R_e$ is typically $0.1 - 0.3$ (see their
Figure 5). In what follows, we adopt $\beta = 0.2 \pm 0.2$ as a
fiducial value, which gives $a=8.05 \pm 2.40$ and $b=1.07 \pm 0.06$.
Also, according to the detailed photometric studies of early-type galaxies
with {\it HST} by Faber et al. (1997) and Ravindranath et al. (2001),
there exists a rather tight, linear correlation between the logarithms of
a central velocity dispersion and a core size in bright galaxies.
Our regression analysis of Faber et al.'s data for cored E/S0 galaxies
(excluding NGC~4486B with a double nucleus) yields
$\log r_c (h_{0.8}^{-1}~{\rm pc}) = 2.62 \log \sigma ({\rm km~s}^{-1})
- 4.01$, where $h_{0.8} \equiv H_0 / 80$ km~s$^{-1}$~Mpc$^{-1}$ and a 1~$\sigma$
error in the determination of $\log r_c ({\rm pc})$ from this
relation is 0.42. Thus, we obtain
$r_c = 120^{+196}_{-74}$ $h^{-1}$~pc at the measured velocity
dispersion of 230 km s$^{-1}$, where $h \equiv H_0 / 100$ km~s$^{-1}$~Mpc$^{-1}$.
In the likelihood analysis described below,
we take into account all of the uncertainties in the above quantities
[$\beta$, $r_c$, $a$, $b$, and $R_e$ in equation (6)] {\it in a consistent
manner} as sources of the uncertainty in the parameter $\sigma_{tot}$.

\subsection{Likelihood analysis}

We perform the maximum likelihood analysis to find the best model
parameters for the reproduction of the current lens system HST~14174$+$5226.
On the assumption that the error in
$\sigma \equiv \langle\sigma_{los}^2\rangle^{1/2}$
is represented by Gaussian distribution, we calculate the likelihood function,
\begin{equation}
L(\Omega_m,\Omega_\Lambda) =
  \frac{1}{\sqrt{2\pi}\delta\sigma}
  \int d\sigma_v \exp \left[ 
  - \frac{(\sigma_v-\sigma)^2}{2\delta\sigma} \right]
  L(\Omega_m,\Omega_\Lambda,\sigma_v) ,
\end{equation}
where
\begin{equation}
L(\Omega_m,\Omega_\Lambda,\sigma_v) \propto
  \exp \left[ - \sum_{i=1}^{4}
  \left( \frac{(\theta_1^{i,model}-\theta_1^{i,obs})^2}
              {2\delta{\theta_1^{i,obs}}^2}
  + \frac{(\theta_2^{i,model}-\theta_2^{i,obs})^2}
              {2\delta{\theta_2^{i,obs}}^2}
  \right) \right] .
\end{equation}
In the above equations, $\delta\theta^{i,obs}$ ($i=1$ ... 4) and
$\delta\sigma$ denote the standard errors in the observed image
positions and velocity dispersion, respectively. We then maximize
$L(\Omega_m,\Omega_\Lambda)$ to find the best set of
cosmological parameters $(\Omega_m,\Omega_\Lambda)$. We note that due
to the constraint in equation (6) for $\sigma_{tot}$ and
$\langle\sigma_{los}^2\rangle^{1/2}$, the number of parameters in the
model is 6 for the CIE model and 8 for the CIE+ES and CIE+LP models.
Since we shall restrict our analyses to flat
($\Omega_\Lambda=1-\Omega_{\rm m}$) or open world models
($\Omega_\Lambda=0$), the number of parameters is further reduced to
5 for the CIE model and 7 for the CIE+ES and CIE+LP models.

\subsection{Results}

Figure 4 shows the best-fit image positions (solid circles) and source
position (cross), compared with the observed image positions and galaxy
position (open circles), where the radii of the open circles correspond to
1 $\sigma$ observational uncertainties. Table 4 tabulates the list of
the best-fit lens parameters and the corresponding minima of $\chi^2$
divided by the number of degrees of freedom, $N_{dof}$. The best-fit CIE
model (Figure 4a) to the observed images yields
$\chi_{min}^2/N_{dof} = 161/3$, which is not satisfactory from the statistical
point of view. It is
worth noting that both the best-fit lens parameters and the value of
$\chi_{min}^2/N_{dof}$ are similar to those obtained by Keeton et al.
(1998) using a singular isothermal ellipsoid or SIS with an external shear.
The inclusion of an external
shear field (CIE+ES: Figure 4b) does not improve the goodness of the fit
significantly: we obtain $\chi_{min}^2/N_{dof} = 121/1$. In contrast to
these two cases, the dramatic improvement of the fit is attained if the
galaxy center is a free parameter (CIE+LP: Figure 4c), yielding
$\chi_{min}^2/N_{dof} = 13/1$, although the predicted lens position (plus
in panel c) deviates rather largely from the observed position.

Although none of the three different models considered here are statistically
acceptable to reproduce the configuration of the current lens system in a
{\it precise} manner, all models produce a similar mean separation between
the four images, which is also similar to the observed image separation;
our experiments show that the variation of the mean separation among the models
is less than $\sim 0.\arcsec1$ at a fixed set of $(\Omega_m,\Omega_\Lambda)$.
This in turn suggests that the limits on the geometry of the universe,
$(\Omega_m,\Omega_\Lambda)$, are not significantly different in each model.
This is more clearly seen in Figure 5, where we plot the likelihood $L$
normalized by its
maximum value as a function of $\Omega_m$. In each model, the lower and upper
lines show the flat $(\Omega_\Lambda = 1 - \Omega_m)$ and open
$(\Omega_\Lambda=0)$ universes, respectively. As is evident, each model
yields a similar
$L$ vs. $\Omega_m$ relation, showing a large probability at a small
$\Omega_m$ for the flat universe, whereas for the open universe, the
probability stays below that for the flat universe. The largest probability is
attained at $\Omega_m \sim 0$ for all three cases. This may imply
that the observed image configuration and the velocity dispersion of the lens
system HST~14176$+$5226 is best reproduced at a large $\Omega_\Lambda$:
formal 1~$\sigma$ lower limit on $\Omega_\Lambda$ in the flat universe
is about 0.9 and 2~$\sigma$ lower limit reads 0.26, 0.62, and 0.00
for the CIE, CIE+ES, and CIE+LP models, respectively (Table 4).

We further investigate the possible systematic effects on our lens models by
changing the fiducial model parameters $(\beta,r_c)$ and also by accounting for
an additional lensing source. First, a $1~\sigma$ shift of the mean $\beta$
while maintaining the error of $\pm 0.2$ corresponds to the more extremely
anisotropic velocity fields with more tangential ($\beta = 0.0 \pm 0.2$) or
radial ($\beta = 0.4 \pm 0.2$) motions of stars in the central region of the
lens. We note here that the coefficients $(a,b)$ in equation (6) change
accordingly, as ($9.01\pm2.05,1.14\pm0.05$) and ($6.98\pm2.82,1.00\pm0.07$),
respectively. Although such extreme velocity fields are unlikely among local
elliptical galaxies (Gerhard et al. 2001), our experiments suggest that the
effect on the determination of the cosmological parameters is not so
significant: the change in $1~\sigma$ lower limit on $\Omega_\Lambda$ amounts
to only $+0.10$ and $-0.04$ for the CIE model with $\beta = 0.0 \pm 0.2$ and
$\beta = 0.4 \pm 0.2$, respectively (Table 4). It is also worth noting
that $2~\sigma$ lower limit on $\Omega_\Lambda$ is somewhat sensitive to
this systematic effect, yielding no constraint for $\beta = 0.0 \pm 0.2$ and
the change of $-0.66$ for $\beta = 0.4 \pm 0.2$. Second, the decrease of a mean
core radius $r_c$ by $1~\sigma$ [while maintaining the error of $\pm 0.42$
in $\log r_c$ ($h^{-1}$~pc)] yields the increase of lower limit on
$\Omega_\Lambda$ by an amount of 0.04 ($1~\sigma$) and 0.62 ($2~\sigma$),
whereas the $1~\sigma$ increase of $r_c$ yields no allowable $1~\sigma$ and
$2~\sigma$ lower limits on $\Omega_\Lambda$. Third, we consider
an additional uniform sheet of lensing matter to the CIE model (CIE+SHEET
in Table 4). This approximates the effect of a possible intervening matter
on the lens system as provided by a foreground cluster or group of galaxies
(e.g., Turner, Ostriker, \& Gott 1984), although there exists no evidence for
a significant enhancement of galaxies in the {\it HST} image of HST~14176$+$5226
(Ratnatunga et al. 1995; 1999; Knudson et al. 2001). The surface density of
the sheet is characterized by a fixed convergence, $\kappa_s$, in units of
the critical surface density for lensing, so that the total number of model
parameters while restricting the universe to be flat is five.
Taking into account the fact that the lens appears {\it not} to reside in the
central part of a cluster of galaxies and that there is no evidence for
arc-like features in the field of view, $\kappa_s$ is expected to be much
below 1. We take 0.1 as a possibly largest limit for $\kappa_s$, which
approximately corresponds to a central surface density of a rich cluster
(Turner, Ostriker, \& Gott 1984). It follows that the additional uniform sheet
with $\kappa_s=0.1$ to the CIE model yields the decrease of lower limit on
$\Omega_\Lambda$ by an amount of 0.20 ($1~\sigma$) and no allowable $2~\sigma$
limit.

Our experiments thus suggest that the systematics in our lens models affect
the confidence interval for reproducing the observed image configuration and
velocity dispersion of the lens system HST~14176$+$5226, whereby 1~$\sigma$
lower limit on $\Omega_\Lambda$ for the flat universe ranges 0.73 to 0.97,
whereas 2~$\sigma$ lower limit is basically unavailable.
Although this is statistically not justified for concluding the presence of
a large $\Omega_\Lambda$, it is worth noting that in whatever models considered
here, the larger $\Omega_\Lambda$ yields the larger likelihood for reproducing
the current lens system.

%%%%% Sec. 5 %%%%%
\section{Discussion and Concluding Remarks}

The current methodology for determining cosmological parameters is quite
insensitive to both dust absorption and evolutionary correction in a lens
system. The ambiguity associated with the effects of dust absorption, if any,
in the lens concerns the current method in only an indirect manner,
through the calibrations of, e.g., the galaxy center, effective radius, and
surface-brightness weighted average of $\sigma_{los}^2$, where all of the
associated errors are expected to be trivial or well within uncertainties
of mass models. Also, since we have not utilized the lens magnitude in deriving
the velocity dispersion of the lens, our analysis exempts from the ambiguity
associated with the luminosity evolution of the lens galaxy, unlike
Im et al. (1997). We suspect that the ambiguity involved in utilizing the
photometric data is rather significant due to the following considerations.
We adopt the Poggianti (1997) work for {\it k}-corrections and her E and E2
models for evolution of ellipticals with a current age of 15~Gyr,
where e-folding times for star formation rate are 1~Gyr and 1.4~Gyr,
respectively. Then, the available photometric data ($V=21.74$:
Ratnatunga et al. 1999, $B-V=1.4$: Crampton et al. 1996) in conjunction with
the Faber-Jackson relation between central velocity dispersion
$\sigma$ and absolute $B$ magnitude $M_B$, $\sigma = 225$ km~s$^{-1}$
$\times 10^{0.1 (M^{\ast}_B - M_B)}$, where $M^{\ast}_B = -19.9 + 5 \log h$
(Im et al. 1997), yield $\sigma$ of 241 km~s$^{-1}$ (E) and 183 km~s$^{-1}$
(E2) for the cosmological parameters of
$(\Omega_m,\Omega_\Lambda,h)=(1,0,0.43)$ (giving the age of the universe
of 15~Gyr). Thus, the velocity dispersion derived in this manner is sensitive
to galaxy evolution models and its possible range is well beyond the error of
the spectroscopic measurements.

We also note that the critical assumption behind our analysis is that
the lens galaxy at $z_L=0.81$ is already in dynamical equilibrium.
This appears to be well guaranteed, because various lines of observational
evidence, including the analyses of the Mg indices vs. $\sigma$ relation
(Ziegler \& Bender 1997; Bernardi et al. 1998) and fundamental plane
(van~Dokkum et al. 1998) of remote ellipticals, have revealed a much higher
redshift for the formation epoch of elliptical galaxies,
typically at $z \ga 3$.

The principal source of ambiguities in the current methodology is
the variety of possible mass models which equally fit to a lens system.
Although the relative significance of this effect is partially diminished
by carefully selecting
the preferable combination of $(z_L, z_S)$ so as to make the distance ratio
$D_{ls}/D_s$ being sensitive to $\Omega_\Lambda$ (as demonstrated
in Figure 1), the large variety in mass models remains unavoidable.
Within the range of our mass models, $1~\sigma$ lower limit on $\Omega_\Lambda$
varies about 0.2 and there exists a much larger variation in $2~\sigma$ lower
limit. Furthermore, we are unable to set a meaningful upper limit on
$\Omega_\Lambda$.
Our mass models for the lens of HST~14176$+$5226 have an isothermal density
profile ($\rho \propto r^n$ with $n=2$), which is {\it statistically} in good
agreement with general properties of field ellipticals (e.g. Rix et al. 1997;
Gerhard et al. 2001). Although the deviation from $n=2$ appears to be
small in most ellipticals (Gerhard et al. 2001), the smaller (larger) exponent
than 2 yields larger (smaller) $\Omega_\Lambda$, because the total mass of
the lens interior to the location of the background lensed images decreases
(increases).
Effects of non-spherical mass density and/or luminosity density
in stellar dynamics may be most uncertain to quantify, because these have
not been fully studied yet (e.g. Keeton et al. 1997). Although a detailed
examination of the issue is well beyond the scope of this work,
it will be important to explore more extensive modeling of internal dynamics
of elliptical galaxies for the application of lensing studies.

HST~14176$+$5226 is one of the ideal sites for applying the current
methodology, mainly because of the high redshift of the lens galaxy.
In addition to this lens system, in order to minimize the uncertainties
associated with the mass models
and/or the individual nature of the galaxy, it is important to measure
velocity dispersions for more sample lenses at high redshift,
thereby placing a tighter and less uncertain constraint on the world model.
We note that the use of Einstein ring systems may
significantly improve the accuracy of the lens modeling which brings about
a much more severe constraint on $\Omega_\Lambda$.
In this regard, it is worth noting that in the $(\Omega_m,\Omega_\Lambda)$
plane, the current method provides a constraint roughly along the
$\Omega_\Lambda \sim const.$ line in the range $0 \la \Omega_\Lambda \la 1$,
i.e. crossing the constraints given by Type Ia results ($\Omega_\Lambda
\simeq \Omega_m + const.$) and by CMB results ($\Omega_\Lambda
= - \Omega_m + const.$). Thus, along with these other cosmological
observations, the current method also provides useful information on
the combination of $(\Omega_m,\Omega_\Lambda)$.
Also, in addition to the value of $\Omega_\Lambda$ itself, further
observations of various lens systems can be used to derive a useful constraint
on the time variability of $\Omega_\Lambda$ (Futamase \& Yoshida 2001;
Yamamoto \& Futamase 2001) or quintessence (Caldwell et al. 1998),
by employing lenses at various redshifts; to reveal time dependence or
independence of the cosmological constant or dark energy shall offer
a great impact not only on cosmology but also on theoretical physics.

\acknowledgments
We are grateful to all members of the Subaru Telescope, NAOJ, Japan,
for their support of the observations. We are also grateful to many
of our colleagues for useful discussion and comments on this work.
T.H. thanks G.~B\"orner and M.~Bartelmann for the hospitality during his
stay at Max-Planck-Institut f\"ur Astrophysik where a part of the present 
work was performed. He also acknowledges support from Research Fellowships 
of the Japan Society for the Promotion of Science.
Numerical computation in the present work was partly carried out at the 
Yukawa Institute Computer Facility.

\clearpage
%%%%%%%%%%%%%%%%%%%%%%%%%%%%%%%%%%%%%%%%%%%%%%%%%%%%%%%%%%%%%%

%%%%%%%%%%%%%%%%%%%%%%%%%%%%%%%%%%%%%%%%%%%%%%%%%%%%%%%
%%% table 1 %%%%%%%%%%%%%%%%%%%%%%%%%%%%%%%%%%%%%%%%%%%
%%%%%%%%%%%%%%%%%%%%%%%%%%%%%%%%%%%%%%%%%%%%%%%%%%%%%%%

\begin{table}
\tablenum{1}
\begin{center}
\caption{Properties of Stars Used as the Spectral Templates}

\begin{tabular}{lcccc}
\tableline\tableline
Star & Type & $T_{\rm eff}$ & log $g$ & [Fe/H] \nl
(1) & (2) & (3) & (4) & (5) \nl
\hline
HD 83805 & G8III & 5038 & 2.14 & $-0.06$ \nl
HD 8491 & K0III & 4703 & 2.34 & $+0.09$ \nl
HD 94247 & K3III & 4165 & 2.30 & $+0.00$ \nl
\tableline
\end{tabular}
\tablenotetext{1}{Star name}
\tablenotetext{2}{Spectral type}
\tablenotetext{3}{Effective temperature (K)}
\tablenotetext{4}{Surface gravity in log}
\tablenotetext{5}{Metal abundance in units of solar metallicity}
\end{center}
\end{table}

\clearpage

%%%%%%%%%%%%%%%%%%%%%%%%%%%%%%%%%%%%%%%%%%%%%%%%%%%%%%%
%%% table 2 %%%%%%%%%%%%%%%%%%%%%%%%%%%%%%%%%%%%%%%%%%%
%%%%%%%%%%%%%%%%%%%%%%%%%%%%%%%%%%%%%%%%%%%%%%%%%%%%%%%

\begin{deluxetable}{ccccc}
%%\footnotesize
\tablenum{2}
\tablewidth{0pc}
\tablecaption{Results of the Fourier Cross-Correlation}
\tablehead{
\colhead{Star} & \colhead{Fitting region\tablenotemark{a}} &
\colhead{CCF fitting width} & \colhead{$\sigma$\tablenotemark{b}} &
\colhead{$R$\tablenotemark{c}}          \\
\colhead{}     & \colhead{}   &
\colhead{}     & \colhead{(km s$^{-1}$)} &
\colhead{}   }
\startdata
HD 83805 & A & 30 & 221 & 14.5 \nl
          & A & 35 & 230 & 14.4 \nl
          & B & 30 & 225 & 15.4 \nl
          & B & 35 & 232 & 15.3 \nl
HD 8491 & A & 30 & 217 & 13.7 \nl
          & A & 35 & 227 & 13.6 \nl
          & C & 30 & 227 & 14.2 \nl
          & C & 35 & 235 & 14.2 \nl
HD 94247 & A & 30 & 233 & 12.3 \nl
          & A & 35 & 244 & 12.2 \nl
          & C & 30 & 233 & 12.4 \nl
          & C & 35 & 243 & 12.4 \nl
%\hline
$R$-weighted mean & & & 230$\pm$8 & \nl
\enddata
\tablenotetext{a}{A: $4250-4450$\AA, B: $4250-4470$\AA, and C: $4250-4500$\AA .}
\tablenotetext{b}{Measured velocity dispersion.}
\tablenotetext{c}{Tonry-Davis $R$ value.}

\end{deluxetable}

\clearpage

%%%%%%%%%%%%%%%%%%%%%%%%%%%%%%%%%%%%%%%%%%%%%%%%%%%%%%%
%%% table 3 %%%%%%%%%%%%%%%%%%%%%%%%%%%%%%%%%%%%%%%%%%%
%%%%%%%%%%%%%%%%%%%%%%%%%%%%%%%%%%%%%%%%%%%%%%%%%%%%%%%

\begin{center}
\begin{deluxetable}{lc}
%%\footnotesize
\tablenum{3}
\tablewidth{0pc}
\tablecaption{Error Sources in Velocity Dispersion Measurement}
\tablehead{
\colhead{Category} &
\colhead{Uncertainties (1$\sigma$)}
}
\startdata
Velocity matching of spectra\tablenotemark{a} & 4 \nl
CCF fitting error\tablenotemark{b} & 10 \nl
FXCOR parameters\tablenotemark{c} & 8 \nl
\hline
Total & 14\tablenotemark{d} \nl
\enddata
\tablenotetext{a}{Error in matching the velocity resolution between spectra
of templates and that of HST~14176$+$5226.}
\tablenotetext{b}{Error in fitting the cross correlation function
(CCF) peaks with a Gaussian function.}
\tablenotetext{c}{Uncertainties in selecting template, wavelength region, and
fitting width of the CCF peak within FXCOR task.}
\tablenotetext{d}{$(4^2+10^2+8^2)^{0.5}\simeq 14$}
\end{deluxetable}
\end{center}

\clearpage

%%%%%%%%%%%%%%%%%%%%%%%%%%%%%%%%%%%%%%%%%%%%%%%%%%%%%%%
%%% table 4 %%%%%%%%%%%%%%%%%%%%%%%%%%%%%%%%%%%%%%%%%%%
%%%%%%%%%%%%%%%%%%%%%%%%%%%%%%%%%%%%%%%%%%%%%%%%%%%%%%%

\begin{deluxetable}{lcccccccc}
\footnotesize
\tablenum{4}
\tablewidth{0pc}
\tablecaption{Results of Maximum Likelihood Analysis\tablenotemark{a}}
\tablehead{
\colhead{Model}              &
\colhead{$\epsilon$}         & 
\colhead{$\phi_\epsilon$}    &
\colhead{$\gamma$}           & 
\colhead{$\phi_\gamma$}      &
\colhead{$\delta x$\tablenotemark{b}}          & 
\colhead{$\delta y$\tablenotemark{c}}          &       
\colhead{$\chi_{min}^2/{\rm N}_{dof}$\tablenotemark{d}}   &
\colhead{$\Omega_{\rm m}$\tablenotemark{e}}                           \\
\colhead{}              &
\colhead{}              & 
\colhead{($^\circ$)}    &
\colhead{}              & 
\colhead{($^\circ$)}    &
\colhead{(arcsec)}      & 
\colhead{(arcsec)}      &
\colhead{}              & 
\colhead{}}
\startdata
CIE                        & 0.30 & 43 & 0 (fixed) & --- & 0 (fixed) & 0
(fixed) & 161/3 & $<0.74$ ($<0.07$)\nl
CIE+ES                     & 0.25 & 18 & 0.15 & 42 & 0 (fixed) & 0
(fixed) & 121/1 & $<0.38$ ($<0.04$)\nl
CIE+LP                     & 0.28 & 41 & 0 (fixed) & --- & 0.017 &
0.25 & 12.7/1 & $<1.0$ ($<0.08$)\nl
CIE ($\beta=0.0\pm0.2$)   & 0.29 & 43 & 0 (fixed) & --- & 0 (fixed) &
0 (fixed) & 161/3 & --- ($<0.17$)\nl
CIE ($\beta=0.4\pm0.2$)   & 0.31 & 43 & 0 (fixed) & --- & 0 (fixed) &
0 (fixed) & 161/3 & $<0.28$ ($<0.03$)\nl
CIE ($r_c=46$ $h^{-1}$~pc)   & 0.31 & 47 & 0 (fixed) & --- & 0 (fixed) &
0 (fixed) & 161/3 & $<0.12$ ($<0.03$)\nl
CIE ($r_c=316$ $h^{-1}$~pc)   & 0.29 & 47 & 0 (fixed) & --- & 0 (fixed) &
0 (fixed) & 161/3 & --- (---)\nl
CIE+SHEET\tablenotemark{f} & 0.29 & 47 & 0 (fixed) & --- & 0 (fixed) & 0
(fixed) & 161/3 & --- ($<0.27$) \nl
\enddata
\tablenotetext{a}{Our standard choice of the parameters $(\beta,r_c)$
includes $\beta = 0.2 \pm 0.2$ and $r_c = 120^{+196}_{-74}$ $h^{-1}$~pc.}
\tablenotetext{b}{The difference between the galaxy center and the
lens center in RA direction in units of arcsec.}
\tablenotetext{c}{The difference between the galaxy center and the
lens center in Dec direction in units of arcsec.}
\tablenotetext{d}{Minimum of $\chi^2$ divided by the number of degrees
of freedom.}
\tablenotetext{e}{Upper limit on $\Omega_{\rm m}$ for the flat universe
($\Omega_{\rm m}+\Omega_\Lambda=1$) at 2 $\sigma$ (1 $\sigma$) confidence
level.}
\tablenotetext{f}{A uniform mass sheet with a surface density $\kappa_s=0.1$
is assumed, in addition to the CIE model.}

\end{deluxetable}

\clearpage
%%%%%%%%%%%%%%%%%%%%%%%%%%%%%%%%%%%%%%%%%%%%%%%%%%%%%%%%%%%%%%

\begin{figure}
%\plotone{f1.eps}
\caption{
Predicted line-of-sight velocity dispersion $\sigma$ of a lens in the simple
SIS approximation (equation 1), as a function of $\Omega_m$.
Solid line shows the case of $z_L=0.81$, $z_S=3.40$, and
$\langle\Delta\theta\rangle/2=1.\arcsec5$ corresponding to
the lens system HST~14176$+$5226
employed in this paper, whereas dotted line shows the case of $z_L=0.30$
while other quantities are fixed. For each case,
the upper and lower lines correspond to $\Omega_\Lambda=0$ and
$\Omega_m+\Omega_\Lambda=1$, respectively.
The comparison between solid and dotted lines indicates that for given
$z_S$ and $\langle\Delta\theta\rangle$, the selection of a lens at
higher $z_L$ makes $\sigma$ being sensitive to the value of $\Omega_\Lambda$
for the flat universe.
}
\end{figure}

%\clearpage

\begin{figure}
%\plotone{f2.eps}
\caption{
Spectrum of the lens elliptical galaxy in the ``Einstein Cross'' system,
HST~14176$+$5226. Several absorption-line features are marked.
}
\end{figure}

%\clearpage

\begin{figure}
%\plotone{f3.eps}
\caption{Normalized rest-frame spectrum of HST~14176$+$5226 (histogram)
overlaied by the normalized template spectrum of HD83805 after convolving
with the Gaussian velocity-broadening function of $\sigma = 230$ km s$^{-1}$
(solid line). The residual spectrum (galaxy spectrum $-$ broadened stellar
template) is also shown. To show how remaining sky spectrum may possibly
affect the fitting, we also show below the sky spectrum which is red-shifted
by the same amount as for the galaxy spectrum.
}
\end{figure}

%\clearpage

\begin{figure}
%\plotone{f4.eps}
\caption{
The lens configuration of the ``Einstein Cross'', HST~14176$+$5226,
compared with the model predictions, (a) CIE, (b) CIE+ES, and (c) CIE+LP.
Open circles show the observed image positions and lens galaxy position,
where the radii of the circles correspond to 1 $\sigma$ observational
uncertainties. The best-fit image positions and source position are
given by solid circles and cross, respectively. In panel (c) for CIE+LP,
the position of the lens galaxy is a free parameter, shown by a plus.
}
\end{figure}

%\clearpage

\begin{figure}
%\plotone{f5.eps}
\caption{
The Likelihood normalized by its maximum value as a function of $\Omega_m$.
Solid, dotted, and dashed lines denote the CIE, CIE+LP, and CIE+ES models,
respectively. In each model, the lower and upper lines show the flat
$(\Omega_\Lambda = 1 - \Omega_m)$ and open world models
$(\Omega_\Lambda=0)$, respectively.
}
\end{figure}

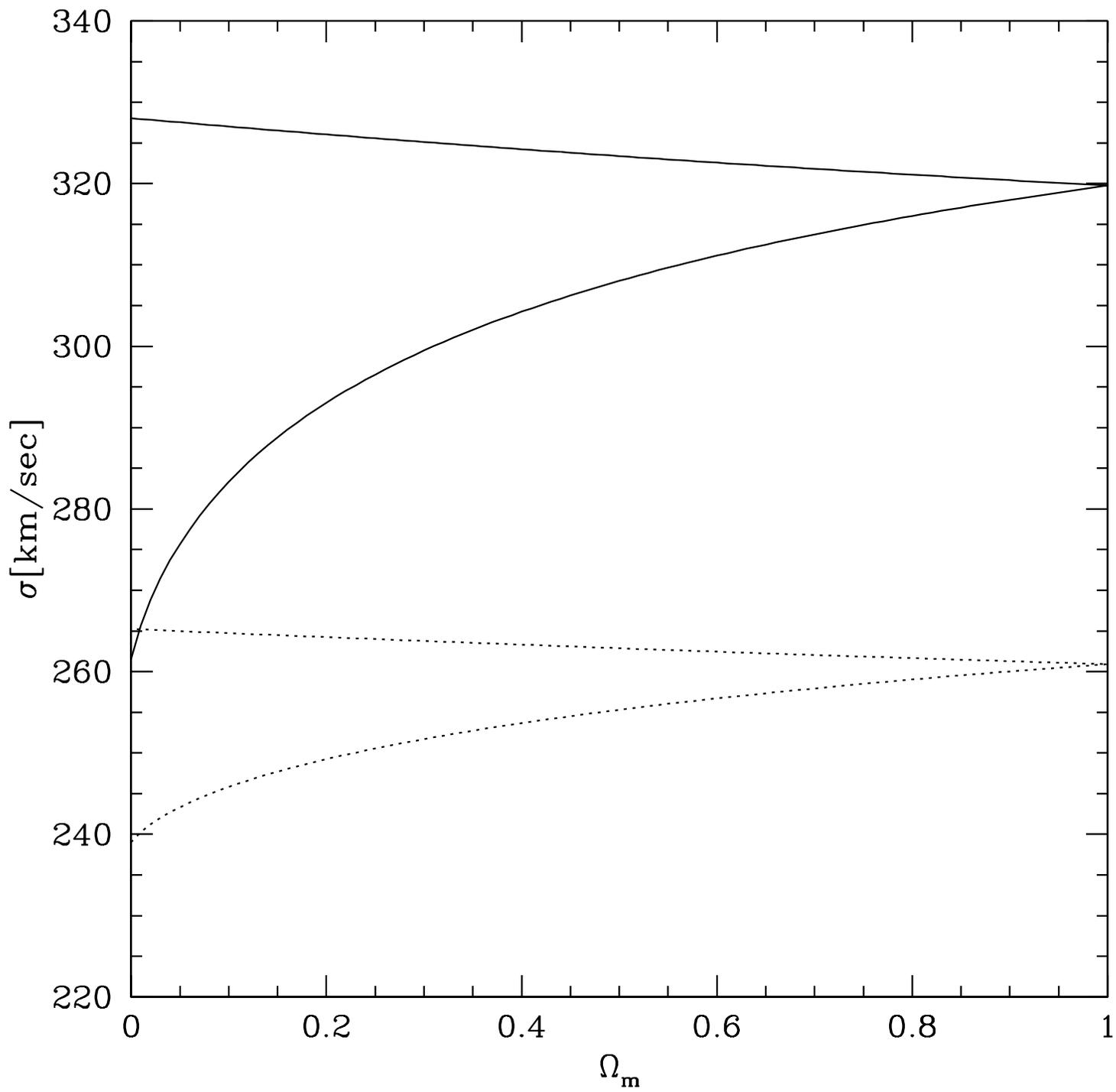
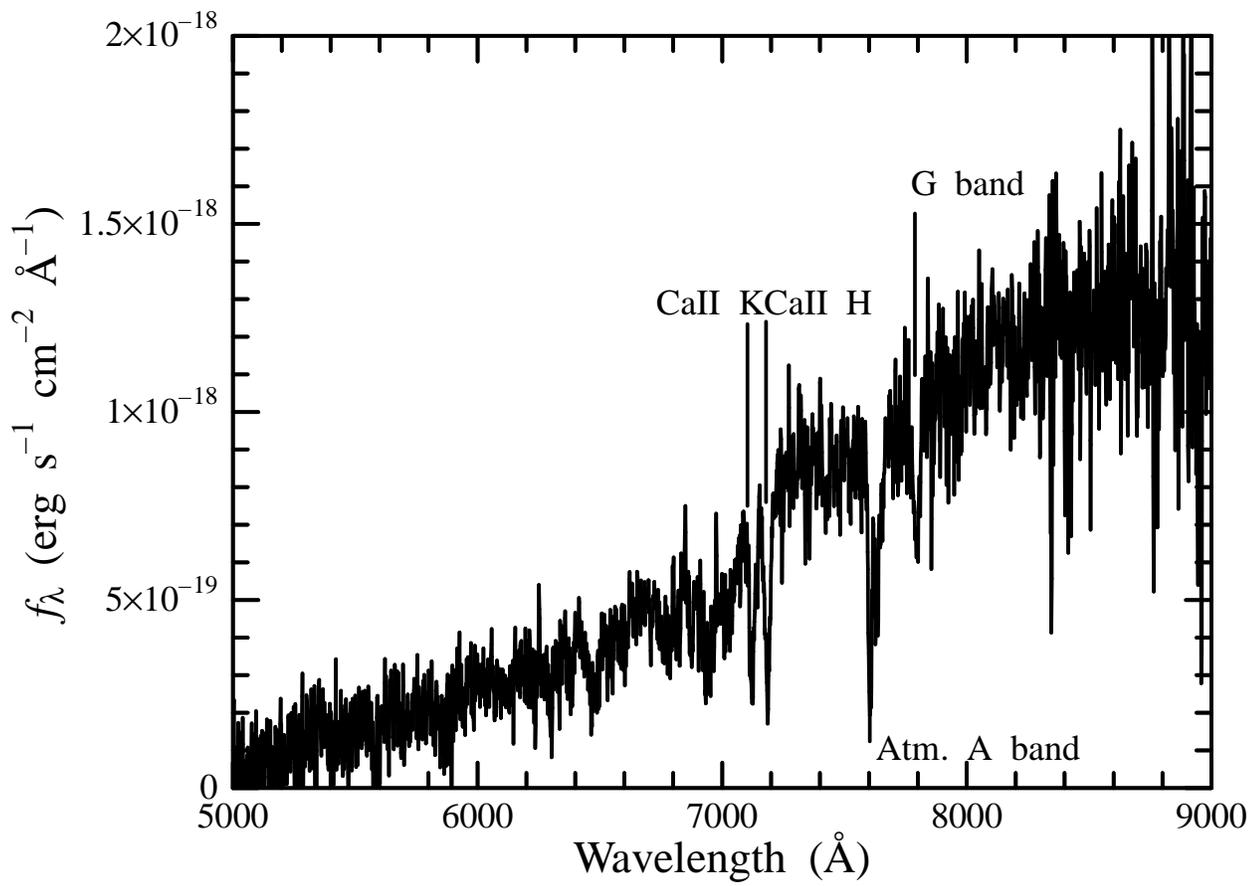
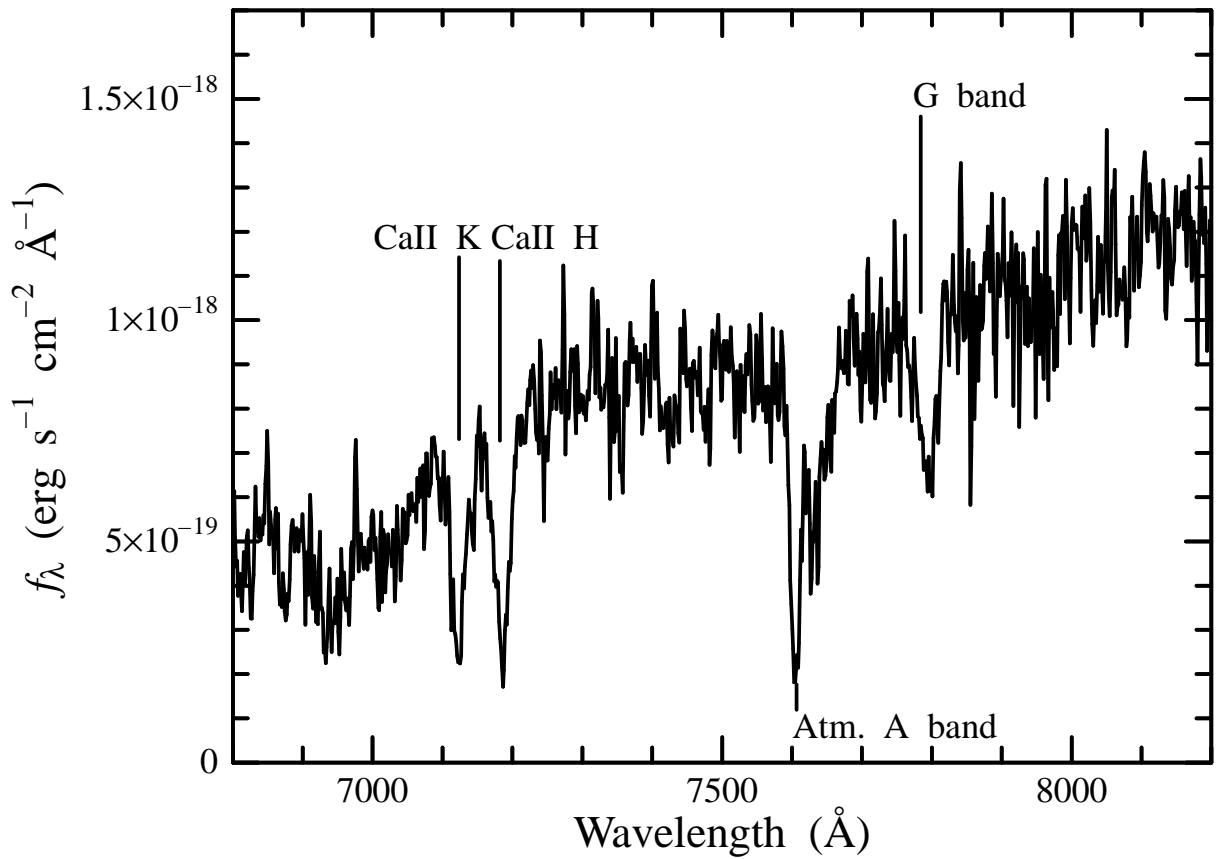
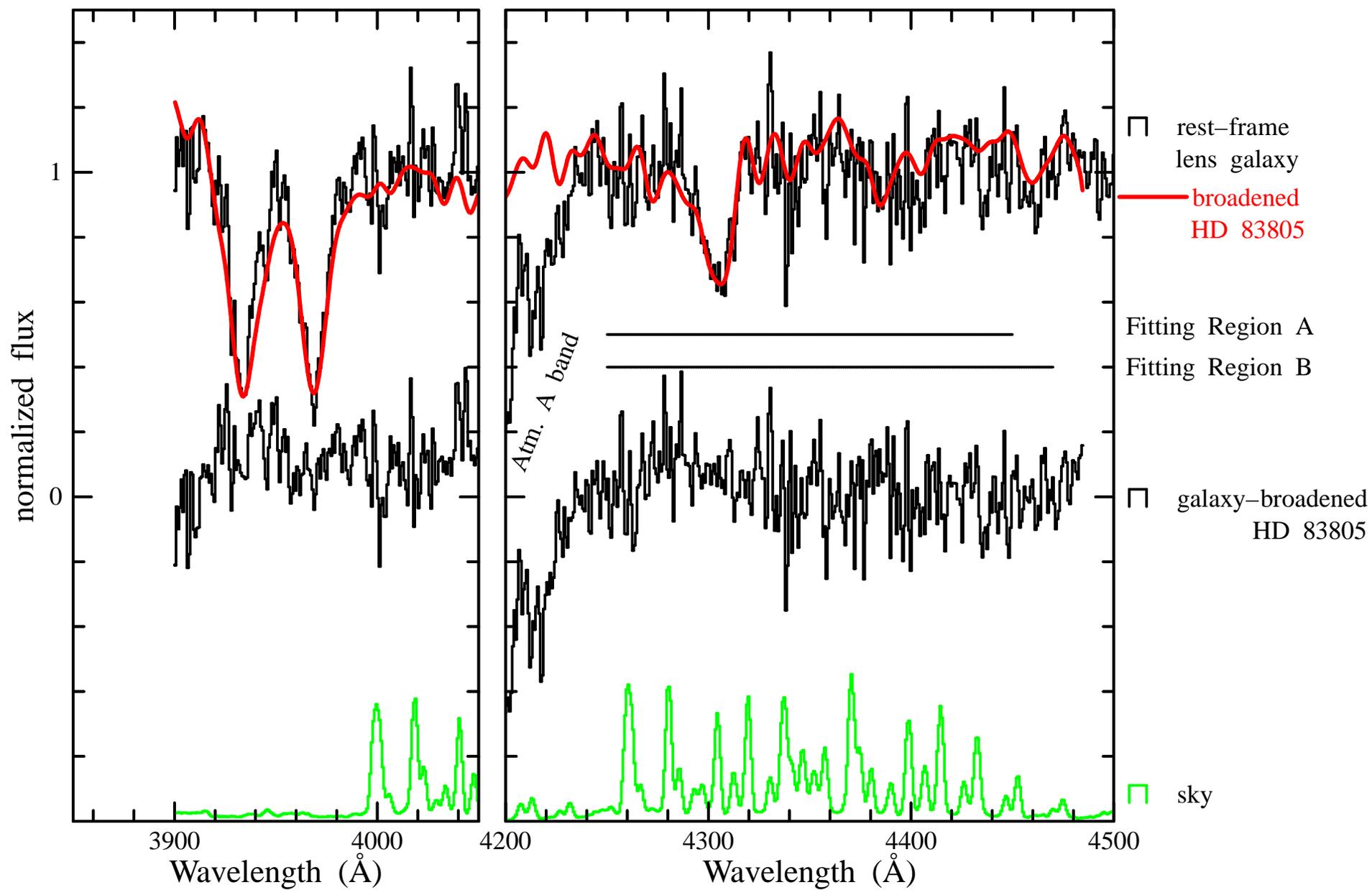
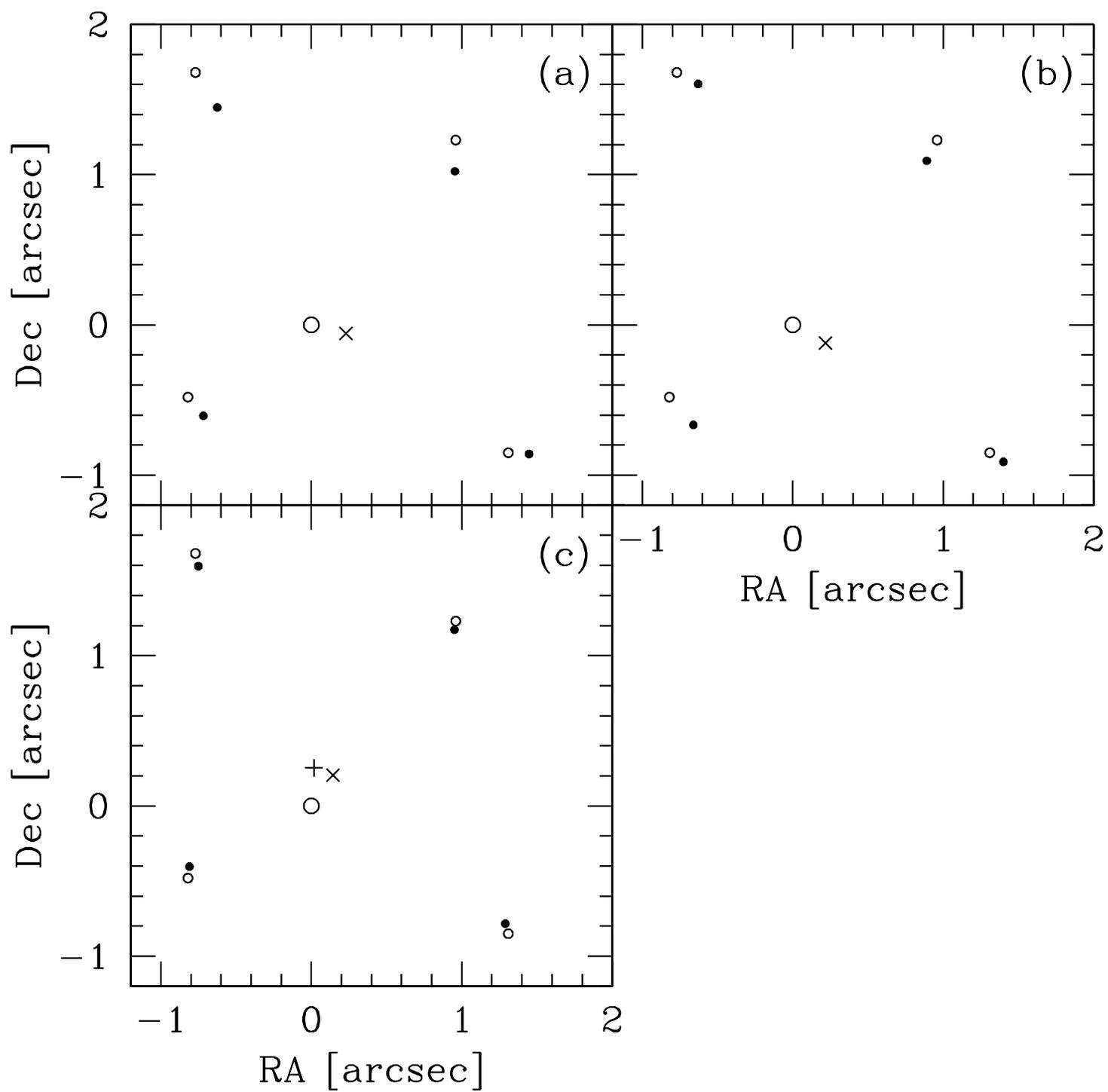
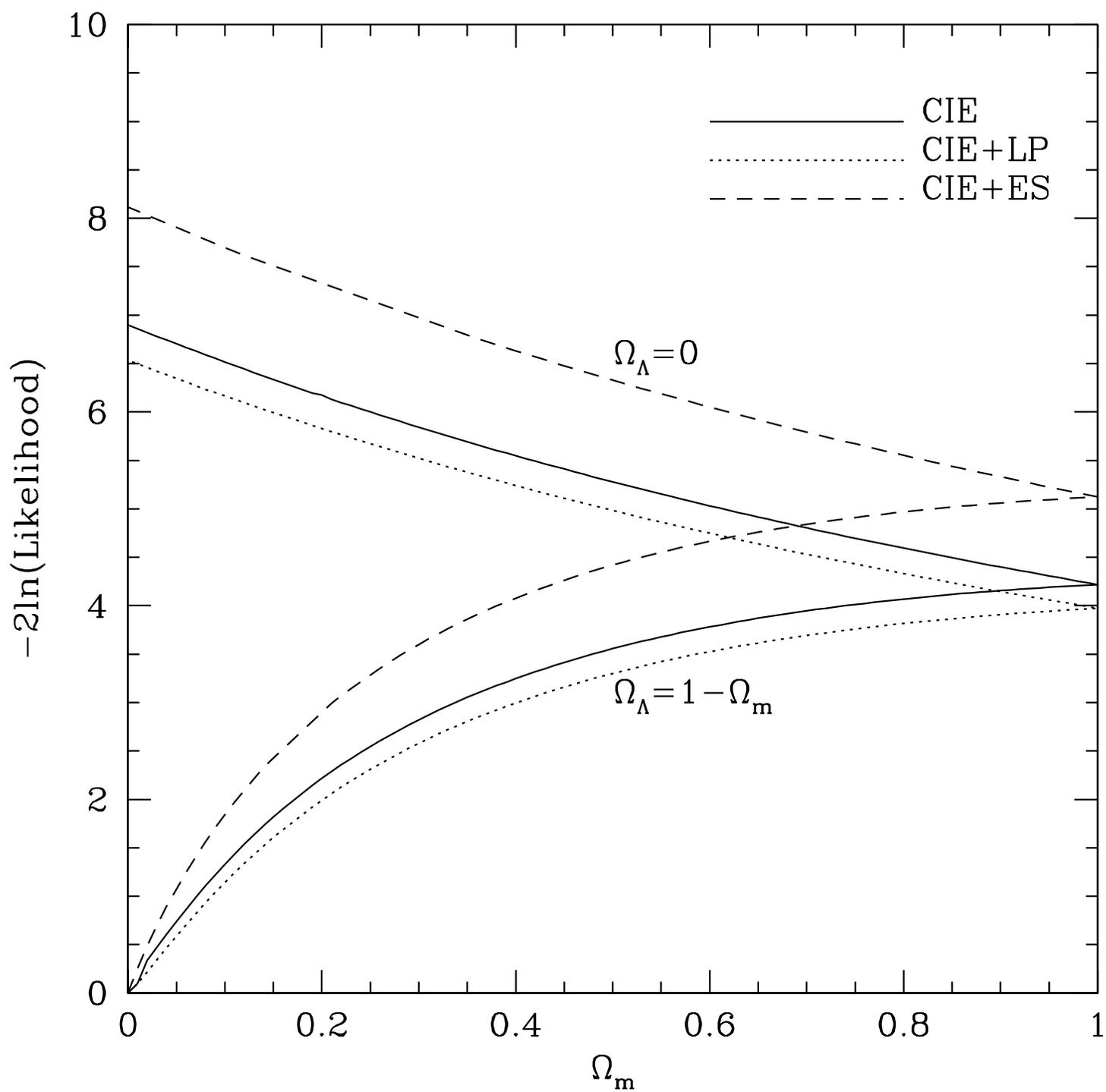
\end{document}